\documentclass[]{spie}

\usepackage{amsmath,amsfonts,amssymb}
\usepackage{graphicx}
\usepackage[colorlinks=true, allcolors=blue]{hyperref}
\usepackage{amsmath}

\usepackage[dvipsnames]{xcolor}

\title{A new window in time: a mid-infrared slit spectroscopy mode for precision time-series astronomy with JWST/MIRI}

\author[a,*]{Achr\`ene Dyrek}
\author[b]{Taylor J.\ Bell}
\author[c]{Pierre-Olivier Lagage}
\author[d]{Sarah Kendrew}
\author[e,f]{Gareb Fern\'andez-Rodr\'iguez}
\author[g]{Thomas Greene}
\author[h]{Giuseppe Morello}
\author[i]{Michiel Min}
\author[b]{Maël Voyer}
\author[f,e]{Hannu Parviainen}

\affil[a]{Space Telescope Science Institute, 3700 San Martin Drive, Baltimore, MD 21218, USA}
\affil[b]{AURA for the European Space Agency (ESA), Space Telescope Science Institute, 3700 San Martin Drive, Baltimore, MD 21218}
\affil[c]{Université Paris-Saclay, Université Paris Cité, CEA, CNRS, AIM, Gif-sur-Yvette, F-91191, France}
\affil[d]{European Space Agency, Space Telescope Science Institute, 3700 San Martin Dr., Baltimore, MD 21218, USA}
\affil[e]{Instituto de Astrof\'isica de Canarias (IAC), 38205 La Laguna, Tenerife, Spain}
\affil[f]{Departamento de Astrofísica, Universidad de La Laguna (ULL), 38206 La Laguna, Tenerife, Spain}
\affil[g]{IPAC, Mail Code 100-22, Caltech, 1200 East California Blvd., Pasadena, CA 91125, USA}
\affil[h]{Instituto de Astrof\'isica de Andaluc\'ia (IAA-CSIC), Glorieta de la Astronom\'ia s/n, 18008 Granada, Spain}
\affil[i]{SRON Space Research Organisation Netherlands, Niels Bohrweg 4, 2333 CA Leiden, The Netherlands}
\authorinfo{Further author information: adyrek@stsci.edu}

\begin{document}
\maketitle

\begin{abstract}
The Mid-Infrared Instrument (MIRI) on board the James Webb Space Telescope (JWST) provides Low Resolution Spectroscopy (LRS) over 5--12~$\mu$m at a resolving power of $R \sim 100$. To date, all MIRI LRS time-series observations (TSOs) have been carried out in slitless mode, since long-duration pointing stability within the narrow $4.7'' \times 0.51''$ slit could not previously be guaranteed, leading to the possibility of degraded TSOs due to jitter. Commissioning activities established the telescope jitter to be less than 1~milliarcsecond (mas), four times less than the initial requirement. It was, therefore, worth assessing the suitability of the MIRI LRS slit as a TSO mode for precision time-domain science; this is the aim of the Cycle~3 program (PID 6219, PI: A.~Dyrek). We observed a transit of the exoplanet HAT-P-12b over $\sim$10~hours and compared the results to archival slitless observations of the same target (PID~1281, PI: P.-O. Lagage). Two independent data reductions of the transit spectra of both slit and slitless configurations are consistent within $1\sigma$, validating the feasibility of the new mode. A joint fit of the slit and slitless observations confirms the presence of a spectral feature near 7.5~$\mu$m. Using measured JWST pointing variations, we estimated slit-loss variations to be smaller than 40 ppm at 10~$\mu$m. The drawback of the slitless mode is a higher background. Compellingly, the slit background is $\sim38$ times lower on average than in slitless mode, improving sensitivity for faint targets ($J_{\rm mag}\sim13$--15). We also identified time-correlated noise unique to the slit dataset at long wavelengths, which requires further investigation. This new capability of TSOs in slit mode, which will be supported in Cycle~7 (in 2028), opens a new avenue for precision time-series astronomy for faint targets with JWST/MIRI.
\end{abstract}


\section{INTRODUCTION}
\label{section:intro}

The Mid-Infrared Instrument (MIRI)\cite{kendrew_mid-infrared_2015} on board the James Webb Space Telescope (JWST) offers a Low Resolution Spectrometer (LRS) mode covering 5--12~$\mu$m at a resolving power of $R \sim 100$. This mode can be operated either in slitless configuration or through a $4.7'' \times 0.51''$ slit (an area of 210 pixels). Time-Series Observations (TSOs) with MIRI LRS require exceptional photometric and pointing stability over durations of many hours in order to detect the small flux variations like those associated with exoplanet transits and eclipses. To date, only the LRS slitless configuration has been offered to the community for low-resolution mid-infrared spectroscopic TSOs \cite{kendrew_mid-infrared_2015,bouwman_spectroscopic_2023}. The telescope's pointing stability over long durations was uncertain, and the risk of introducing complex time-varying detector systematics or slit losses was considered too high. However, commissioning activities demonstrated that JWST is an extremely stable observatory, with pointing variations of only $\sim$1~milliarcsecond (mas)\cite{rigby_science_2023}. The quoted JWST jitter is an RMS line-of-sight pointing stability measurement derived from high-cadence centroid tracking over short (few-minute) intervals. This exceptional stability motivated a GO Cycle~3 CAL program (PID~6219, PI: A. Dyrek) to test whether the MIRI LRS slit could, in fact, be used reliably for precision time-series observations. The use of the slit would offer two key advantages over the slitless mode: (i) significantly reduced background and contamination from nearby sources, which is particularly beneficial for faint targets, and (ii) potentially improved control over detector systematics. The goal of this work is to assess the performance of this new LRS slit TSO mode and to validate its suitability for precision time-domain science.

\section{METHODS \& RESULTS}
\label{section:methods}

\subsection{Detector configuration}

\begin{figure}[htbp]
\centering
\includegraphics[width=0.6\textwidth]{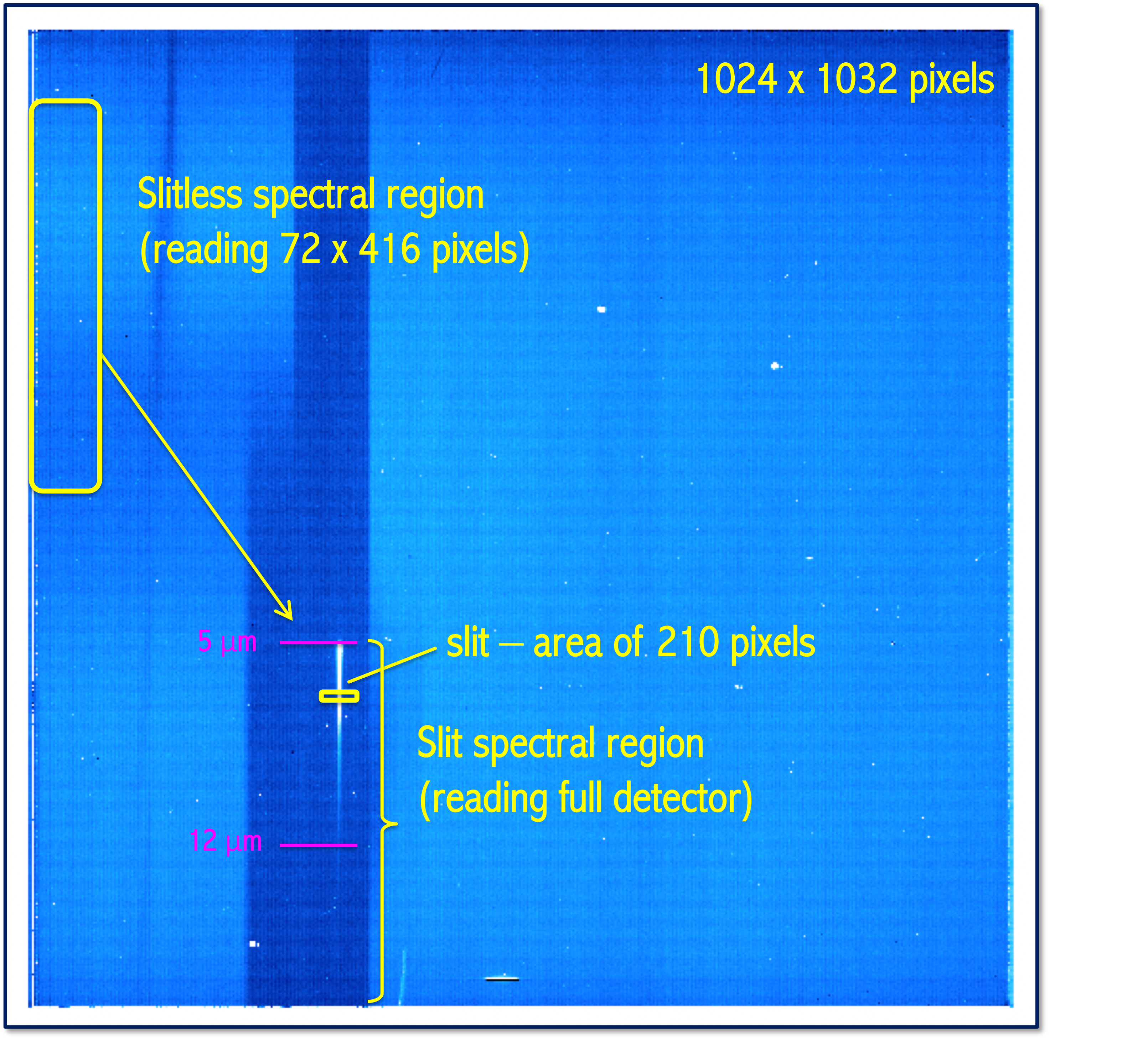}
\caption{MIRI detector showing the regions used for slitless and slit spectroscopy. The slitless spectral region reads out $72 \times 416$ pixels, while the slit spectral region requires reading the full $1024 \times 1032$ pixel detector for the time being; the slit itself corresponds to an area of 210 pixels.}
\label{fig:detector}
\end{figure}

The MIRI detector is constituted of $1024 \times 1032$ pixels. Fig.~\ref{fig:detector} compares the detector regions used for slitless and slit spectroscopy. In slitless mode, only a subarray of $72 \times 416$ pixels is read out. In slit mode, the full detector is read out for the time being, and a smaller region of $45 \times 389$ pixels needs to be extracted by the user.

\subsection{Observational validation}

\begin{figure}[htbp]
\centering
\includegraphics[width=0.7\textwidth]{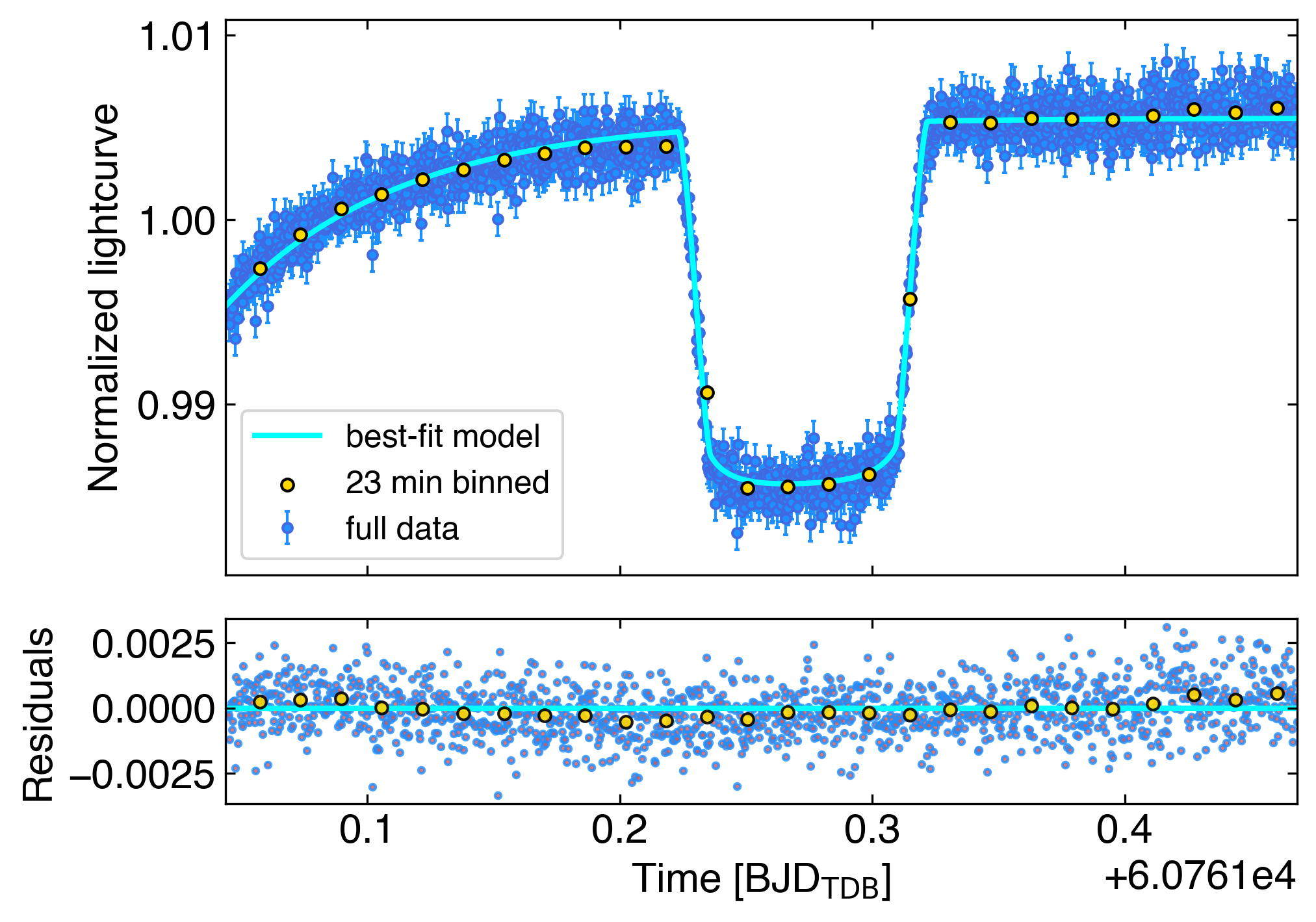}
\caption{Transit observation of the exoplanet HAT-P-12b over $\sim$6~hours, representing the first time-series observation acquired using the MIRI LRS slit, made using the \texttt{Eureka!} pipeline\cite{bell_eureka_2022}.}
\label{fig:white_lightcurve}
\end{figure}

To assess the feasibility of the LRS slit TSO mode, we observed a transit of the exoplanet HAT-P-12b (Fig.~\ref{fig:white_lightcurve}) as part of PID~6215. This target has been extensively studied with JWST\cite{crouzet_detection_2025, heinke_information_2026}. It was observed with MIRI/LRS (slitless) in June 2023, with NIRSpec/PRISM in February 2023 and with NIRISS/SOSS in June 2023, as part of PID~1281. The host star has a magnitude of K=10.1, which makes it well-suited for an LRS slit observation without saturation. Our observational mode was slit spectroscopy using the FASTR1 readout pattern and reading the full detector. HAT-P-12b was observed from March 27 2025 00:54:32 UTC to March 27 2025 11:07:48 UTC with a total duration of 10.22 hours and a transit duration of 2.33 hours. The data was split into 22 uncalibrated file segments with a total of 1326 integrations of 9 groups each. Data reduction was carried out with two independent pipelines as described below. 

\subsubsection{Stage 1 Data Processing}\label{section:stage1}

Early data reductions indicated that these new LRS slit observations were substantially impacted by MIRI's Brighter-Fatter Effect (BFE)\cite{argyriou_brighter-fatter_2023}, where pixels near the centre of the spectral trace (especially at short wavelengths) showed a negative deviation from a linear slope in the up-the-ramp samples, while adjacent pixels showed a positive deviation. As such, we performed Stage 1 data processing using a custom Jupyter notebook developed by Jeroen Bouwman. This notebook first used the \texttt{jwst} pipeline\cite{bushouse_jwst_2025} (version 1.20.2, with CRDS context jwst\_1464.pmap) to run the steps \texttt{dq\_init}, \texttt{emicorr}, \texttt{saturation}, \texttt{firstframe}, \texttt{lastframe}, \texttt{reset}, and \texttt{linearity} steps with default settings. Then the notebook performed a custom BFE correction based on the approach described by ref.\cite{Coulton2018} which involved an iterative deconvolution approach where each observed frame was used as an initial guess of the underlying scene, and a 3$\times$3 Sobel deconvolution kernel was iteratively applied to reconstruct the true underlying scene until a convergence criterion had been reached. After this custom BFE-correction step, deviations in the up-the-ramp samples from a linear trend were reduced by roughly an order of magnitude in the core of the PSF. Stage 1 processing then proceeded with the \texttt{jwst} pipeline's \texttt{dark\_current}, \texttt{rscd}, \texttt{jump} (with the rejection threshold raised to 7.0), and \texttt{ramp\_fit} steps. 

\subsubsection{Eureka! pipeline}\label{section:eureka}

Our fiducial data reduction used the open-source \texttt{Eureka!}\ data analysis pipeline\cite{bell_eureka_2022} (version 1.2.2.dev106+g817c04a2). In Stage 2, we applied the \texttt{assign\_wcs} step but skipped the \texttt{flat\_field}, \texttt{photom}, and \texttt{extract\_1d} steps. In Stage 3, we then cropped the FULL frame to a smaller subarray around the region of the detector corresponding to the LRS slit, masked values marked as DO\_NOT\_USE in the data quality array, and performed two rounds of 5$\sigma$ outlier rejection of the background pixels along each pixel's time-axis. We then determined the source position in each integration by fitting a Gaussian to the 1-D spatial profile computed by collapsing each frame along the wavelength direction using a median; this later allowed us to detrend against changes in the spatial position and width of the PSF in \texttt{Eureka!}'s Stage 5. We then ran two different reductions, one with \texttt{Eureka!}'s standard two-column background subtraction (using the pixels 11-22 pixels away from the centre of the trace), and one reduction without any background subtraction. Our extracted background level was around $\sim$10 DN s$^{-1}$, which is 38$\times$ lower on average than the background when observing with the slitless subarray (see Sect.~\ref{section:background}). Our reductions showed that subtracting the background may have resulted in self-subtraction at the longest wavelengths, biasing the transmission spectrum towards larger values. As a result, we decided not to remove the background in our fiducial reduction. To extract the source flux, we used optimal spectral extraction\cite{horne1986optspec} using the pixels within 5 pix from the centre of the spectral trace and using a cleaned median integration to construct our extraction profile.

In Stage 4, we first masked 3 outlier wavelengths which had higher noise levels compared to nearby wavelengths. We then spectrally binned our data using the same binning scheme used by ref.\cite{heinke_information_2026} for their MIRI/LRS slitless spectrum. They used 36 bins spanning 4.593--11.985 $\mu$m with bin widths ranging from 0.144 to 0.251 $\mu$m and an effective resolving power, $R=\lambda/\Delta\lambda$, generally increasing toward longer wavelengths, from approximately 22 to 59. We also explored the sensitivity of the retrieved spectrum to a different bin size of 0.4 $\mu$m. Outliers in each spectrally-binned lightcurve were then masked using a running median sigma-clipping method with a rejection threshold of 4$\sigma$ and a box size of 20 integrations.

To better constrain the planet's orbital parameters, we first used \texttt{Eureka!} to produce broadband lightcurves from all available JWST datasets of HAT-P-12b which consisted of the two MIRI LRS datasets (5--12 $\mu$m), the NIRSpec/PRISM observation (2.735--3.71 $\mu$m), and the NIRISS/SOSS observation (0.85--2.8 $\mu$m)\cite{heinke_information_2026, crouzet_detection_2025}. We then jointly fit the broadband lightcurves with a shared set of orbital parameters and independent systematic parameters for each instrument. We fitted for the orbital period $P$, the normalised semi-major axis $a/\mathrm{R}_{\star}$, the inclination $i$, and the transit mid-time $t_0$ using Normal priors based on values reported by ref.\cite{Kokori2023} while fixing the orbital eccentricity to 0 and using weakly informative $\mathcal{U}(-1,1)$ priors on the two quadratic limb darkening parameters. Our fits used the \texttt{batman} \cite{kreidberg_batman_2015} transit model. For each instrument, our systematics model included a linear trend in time, an exponential ramp in time, a linear decorrelation against changes in the spatial position and width of the PSF. We also included a white-noise multiplier to account for any additional white noise beyond \texttt{Eureka!}'s estimated stellar-photon-limited noise, and we used a celerite2\cite{celerite1, celerite2} Gaussian process with a Matérn-3/2 kernel to marginalize over the impact of any residual red noise. We also removed the first 50 integrations from the MIRI/LRS datasets and the first 5 integrations from the NIRSpec/PRISM and NIRISS/SOSS datasets to remove the worst of the initial settling trends. We used the nested sampling Bayesian inference framework with the \texttt{dynesty} sampler\cite{speagle_dynesty_2020}, with 2048 live points, `multi' bounds, the `rwalk' sampler, and a convergence criterion of $d\log{\mathcal{Z}}<0.01$, where $\mathcal{Z}$ is the Bayesian evidence. The parameters retrieved from the fit are presented in Table~\ref{table:white_parameters}.

For our spectroscopic fits, we then fixed our orbital parameters to the median values from our broadband fits. An initial fit to the spectroscopic lightcurves with weakly constrained quadratic limb darkening coefficients indicated that the data were consistent with the predicted coefficients from the Stagger stellar model\cite{magic2015stagger} as computed by the ExoTiC-LD tool\cite{Grant2024}. As such, we chose to fix our limb-darkening coefficients to those model predictions. The fitted parameters were the planet-to-star radius ratio $R_p/\mathrm{R_{\star}}$, the same systematics model setup as used in the broadband lightcurve fit, and the same sampling setup as used in the broadband lightcurve fit with the exception of reducing the number of live points to 121 given the far smaller dimensionality of these fits. Our resulting transmission spectrum is shown in Fig.~\ref{fig:slit_spectra}). The residuals of the lightcurve fits showed evidence for time-correlated noise, with a correlation lengthscale of $\sim$1 minute, and the amplitude of the correlation increases at shorter wavelengths. Our use of a Gaussian process allowed us to marginalize over the potential impacts of this residual red noise on our astrophysical inferences, resulting in larger transit depth uncertainties where the residual red noise was correlated with the transit signal (e.g., the shortest wavelength channel) and minimal impact on the uncertainties where there was minimal red noise or the red noise was not correlated with the transit signal.

\begin{table*}[h!]
\caption{Parameters estimated from the white light curve analysis of the JWST MIRI LRS  slit observation of HAT-P-12b. The results
are shown from the Eureka! data analysis on a wavelength range between 5 and 12 $\mathrm{\mu}$m.}  
\label{table:white_parameters} 
\begingroup
\renewcommand{\arraystretch}{1.2}
\centering                   
    \begin{tabular*}{\textwidth}{@{\extracolsep{\fill}}lcc}
    \hline\hline              
    Parameter & Prior & Value \\       
    \multicolumn{3}{l}{Stellar parameters\cite{stassun_revised_2019}} \\
     Star radius [$R_{\odot}$] &  $\cdot \cdot \cdot $ & $0.703$ \\
     Star mass [$M_{\odot}$] &  $\cdot \cdot \cdot $ & $0.74$ \\
     Effective temperature [K] &  $\cdot \cdot \cdot $ & $4652$ \\
    \hline
    \multicolumn{3}{l}{Planetary fixed parameters} \\
    Eccentricity & $\cdot \cdot \cdot$ & 0 \\
    \hline
    \multicolumn{3}{l}{Planetary retrieved parameters from joint white-light-curve fit} \\
    Orbital Period [days] & $\mathcal{N}(3.21305762, 0.00000015)$; ref.\cite{Kokori2023} & $3.213057854^{+0.000000052}_{-0.000000047}$ \\
    Mid-Transit time $t_0$ [BJD\_TDB] & $\mathcal{N}(2456851.481119, 0.000060)$; ref.\cite{Kokori2023} & $2456851.481138^{+0.000048}_{-0.000049}$ \\
    Inclination [$^{\circ}$] & $\mathcal{N}(89.0, 0.4)$; ref.\cite{Kokori2023} & $88.967^{+0.050}_{-0.048}$ \\
    Semi-major Axis [$a/R_{\mathrm{\star}}$] & $\mathcal{N}(11.77, 0.21)$; ref.\cite{Kokori2023} & $11.729 \pm 0.022$ \\
    \hline
    \end{tabular*}
\endgroup
\end{table*}

\subsubsection{The ExoTEDRF+ExoIris pipelines}

\begin{figure}[htbp]
\centering
\includegraphics[width=0.8\textwidth]{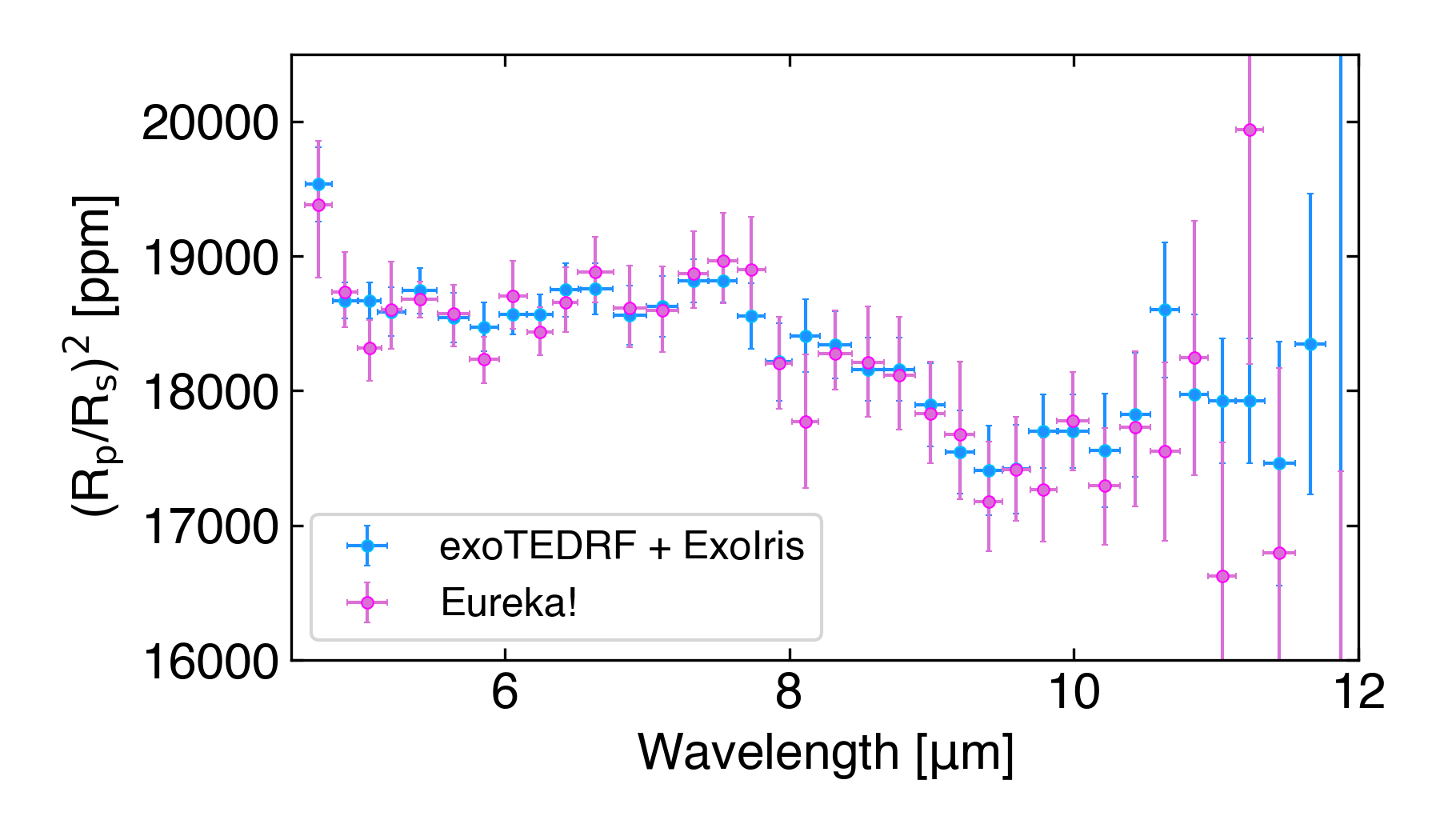}
\caption{Two independent data reductions of the first transit spectrum acquired in the slit mode of MIRI/LRS, showing consistency within $1\sigma$ uncertainties.}
\label{fig:slit_spectra}
\end{figure}

For robustness, we performed a second reduction combining the two pipelines \texttt{ExoTEDRF}\cite{luque_exotedrf_2025,radica_exotedrf_2024,radica_exotedrf_2023,feinstein_exotedrf_2023} and \texttt{ExoIris}\cite{parviainen_exoiris_2026}. We started the reduction at Stage 2 with the outputs from \texttt{Eureka!}'s Stage 1, as described in Sect. \ref{section:eureka}. We skipped flat-field correction and used a 10$\sigma$ threshold for the spatial and temporal bad pixel correction steps. We also skipped the background subtraction, for comparison with the fiducial reduction, and the PCA reconstruction step. In Stage 3, we used the box extraction method with an aperture of 5 pixels. We then binned the data to 0.2~$\mu$m bins. For the equivalent Stages 4--6 of \texttt{Eureka!}\ we used \texttt{ExoIris}. For a thorough explanation of the functionality of the code, please see the corresponding article\cite{parviainen_exoiris_2026}. In summary, unlike regular transmission spectrum extraction methods, which involve fitting light curves individually, \texttt{ExoIris} simultaneously models the 2D spectrophotometric data cube. It also allows joint fitting of multi-epoch and multi-instrument observations. We used a power-2 limb darkening law with wide uninformative uniform priors, normal priors for the transit centre and period, and uniform priors for the radius ratios between 0.1 and 0.2 $R_{\mathrm{p}}/R_{\mathrm{\star}}$. Given the small number of wavelength elements, we used 30 knots and 10 limb darkening knots, so as not to overestimate the wavelength-to-wavelength variation. Moreover, we used the nearest interpolation method. After fitting, we ran 3 runs of MCMC with 3000 iterations each in order to estimate the uncertainties. We fitted in a homogeneous manner the slit, slitless, and joint fit datasets. Due to convergence issues, we restricted the wavelength range to those points below 11~$\mu$m. A comparison between the slit reduction from this pipeline and \texttt{Eureka!} can be seen in Fig.~\ref{fig:slit_spectra}. The slit, slitless, and joint fits are shown in Fig.~\ref{fig:slit_slitless_spectra}.

\subsubsection{Comparison to archival slitless observations}\label{section:slitless}

\begin{figure}[htbp]
\centering
\includegraphics[width=0.8\textwidth]{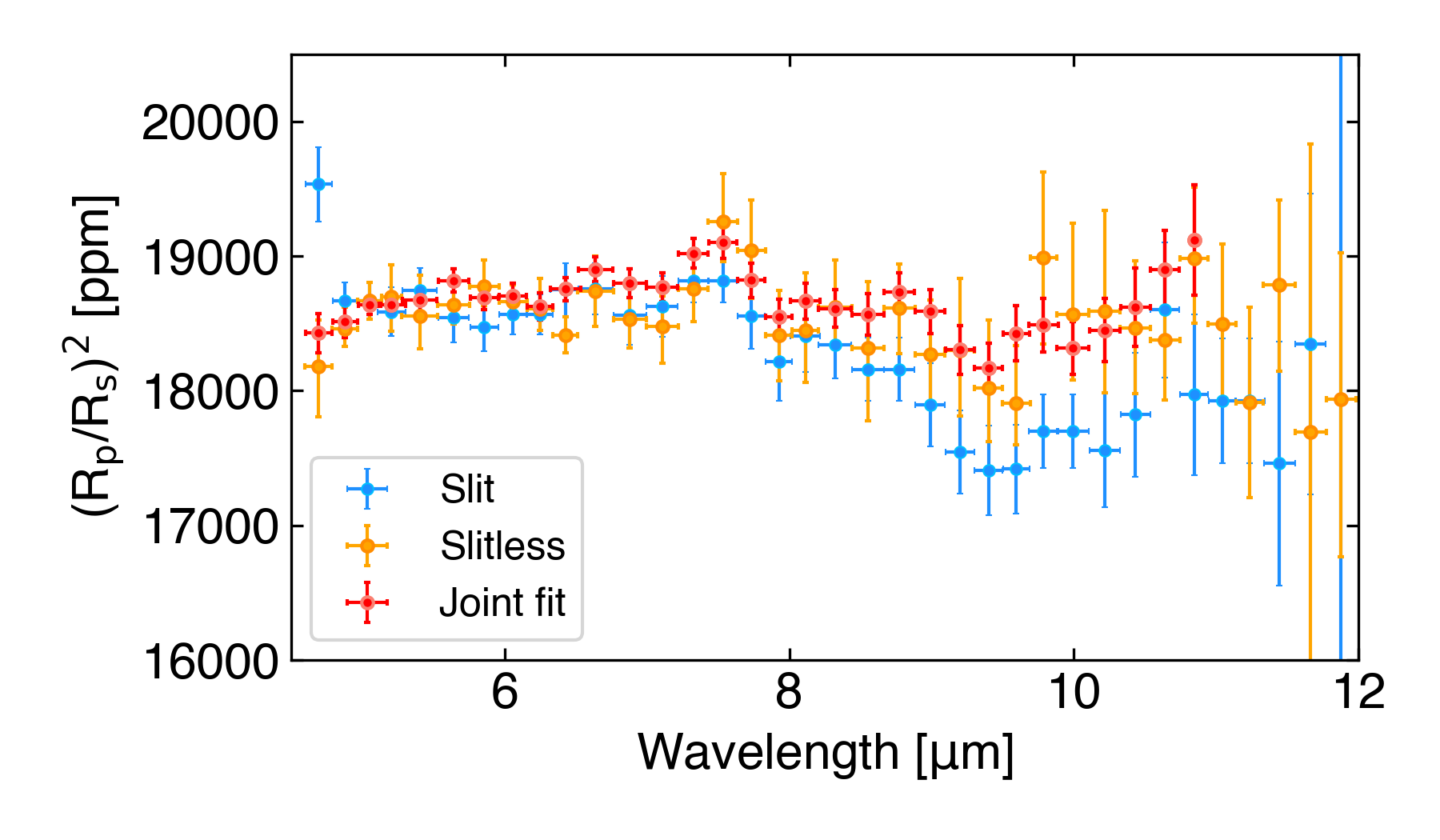}
\caption{Comparison of the transmission spectra of HAT-P-12b retrieved from the MIRI LRS slit (blue) and archival slitless (orange) observations, together with the joint fit of both datasets (red).}
\label{fig:slit_slitless_spectra}
\end{figure}

To validate the performance of the MIRI LRS slit mode, we re-analysed the archival slitless transit observation of HAT-P-12b (PID~1281)\cite{heinke_information_2026}. The slitless dataset was reduced independently with both the \texttt{Eureka!}\ and the \texttt{ExoTEDRF}+\texttt{ExoIris} pipelines, using the same fitting strategy as adopted for the slit observations. In contrast to the slit reduction, background subtraction was performed for the slitless data, as in the slitless configuration each pixel receives the background integrated over the entire wavelength coverage, leading to a significantly larger background contribution in each pixel. We also explored reductions both with and without Gaussian processes to assess the impact of time-correlated noise (see Fig.~\ref{fig:slit_slitless_spectra}). 

We find that there are systematic differences between the slit and slitless spectra. The first spectral channel of the slit spectrum is offset relative to the slitless result, due to residual non-linearity and BFE corrections that are more challenging with only 9 groups per integration, which needed to read out the FULL array, despite the application of a custom BFE correction. The slitless observation was acquired using the dedicated SLITLESSPRISM subarray, providing 150 groups per integration. The substantially larger number of groups in the slitless data enables a more robust ramp fit and improves the correction of detector non-linearities.  At longer wavelengths, the slitless spectrum shows a systematically higher baseline than the slit spectrum. This offset is related to differences in the background subtraction. Finally, we performed a joint fit of the slit and slitless datasets using the \texttt{ExoTEDRF}+\texttt{ExoIris} reduction to assess the consistency of the retrieved transmission spectra. The joint analysis confirms the presence of the absorption feature near 7.5~$\mu$m in both datasets, as was first indicated by ref.\cite{heinke_information_2026}.

Additionally, the spectral dispersion profile of the double prism that is used by the LRS mode folds over on itself around 4 $\mu$m and wavelengths below 4 $\mu$m are dispersed back at longer wavelengths ($\sim$5 $\mu$m). This issue is mitigated when using the slit for TSOs, as a filter is mounted on the slit mask, blocking all radiation below 4.5 $\mu$m.

\subsection{Estimating slit-loss variations}\label{section:slit_loss_variations}

\begin{figure}[htbp]
\centering
\includegraphics[width=0.6\textwidth]{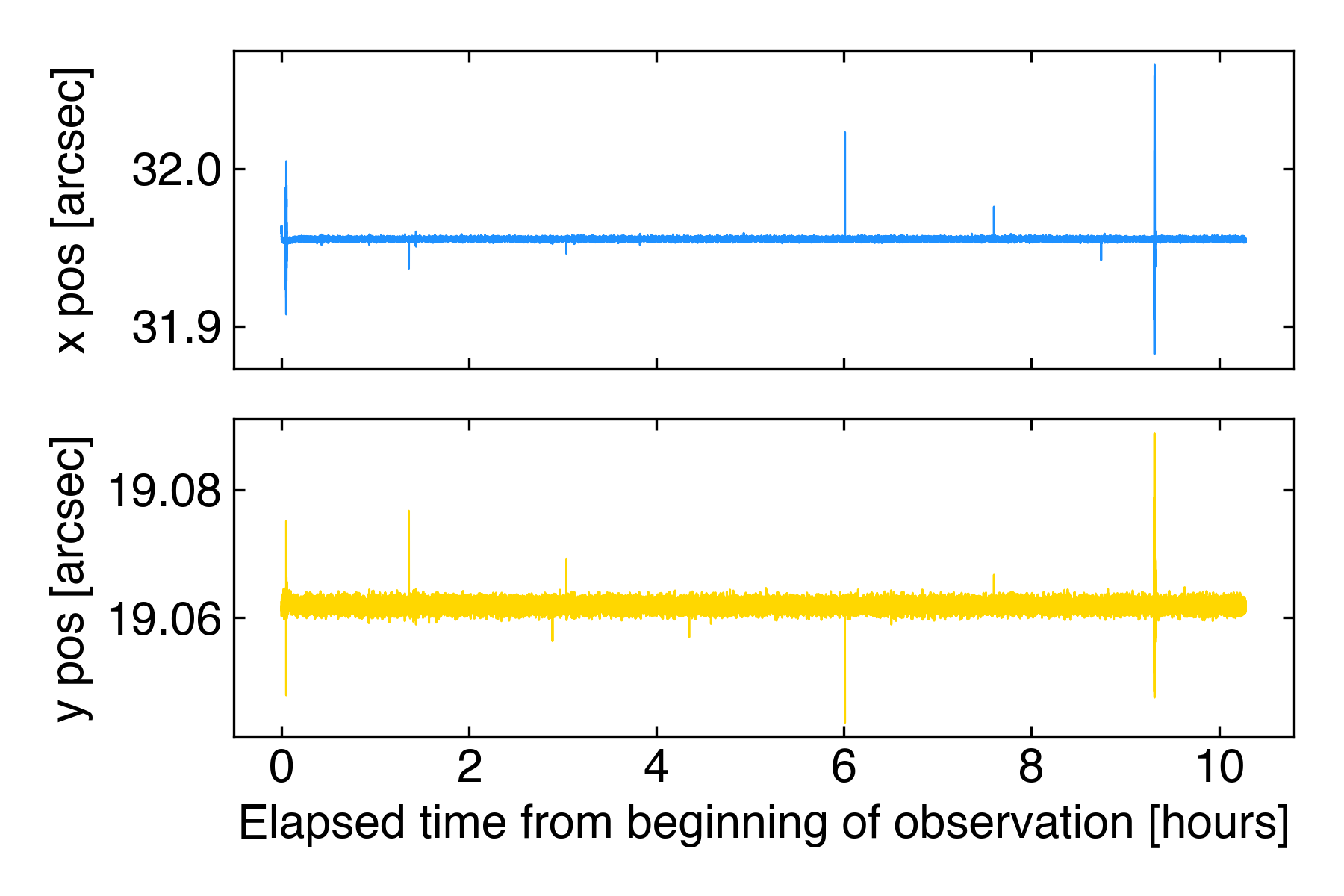}
\caption{Time series of the Fine Guidance Sensor (FGS) guide-star centroid positions during the $\sim$10-hour MIRI LRS slit observation of HAT-P-12b. The top and bottom panels show the centroid positions along the detector $x$ and $y$ directions, respectively.}
\label{fig:fgs}
\end{figure}

A potential limitation of slit spectroscopy for time-series observations is the presence of time-variable slit losses. In fact, the Stage~2 part of the calibration \texttt{jwst} pipeline does include a slit loss correction (the \texttt{path\_loss} step), but this currently assumes the target is perfectly centred in the slit, and it is not currently able to apply a time-variable correction. Slit losses occur when a fraction of the stellar point spread function (PSF) falls outside the slit aperture. Temporal variations of the telescope pointing may, therefore, induce changes in the amount of stellar flux transmitted through the slit, producing systematic photometric variations that could bias the retrieved transmission spectrum. Quantifying the expected amplitude of these variations is therefore essential to assess the suitability of the MIRI/LRS slit mode for high-precision transit spectroscopy. To estimate the expected slit-loss variations, we first measured the pointing stability of the telescope during our $\sim$10-hour observation using the Fine Guidance Sensor (FGS) guide-star centroid telemetry (Fig.~\ref{fig:fgs}). 

\begin{figure}[htbp]
\centering
\includegraphics[width=0.6\textwidth]{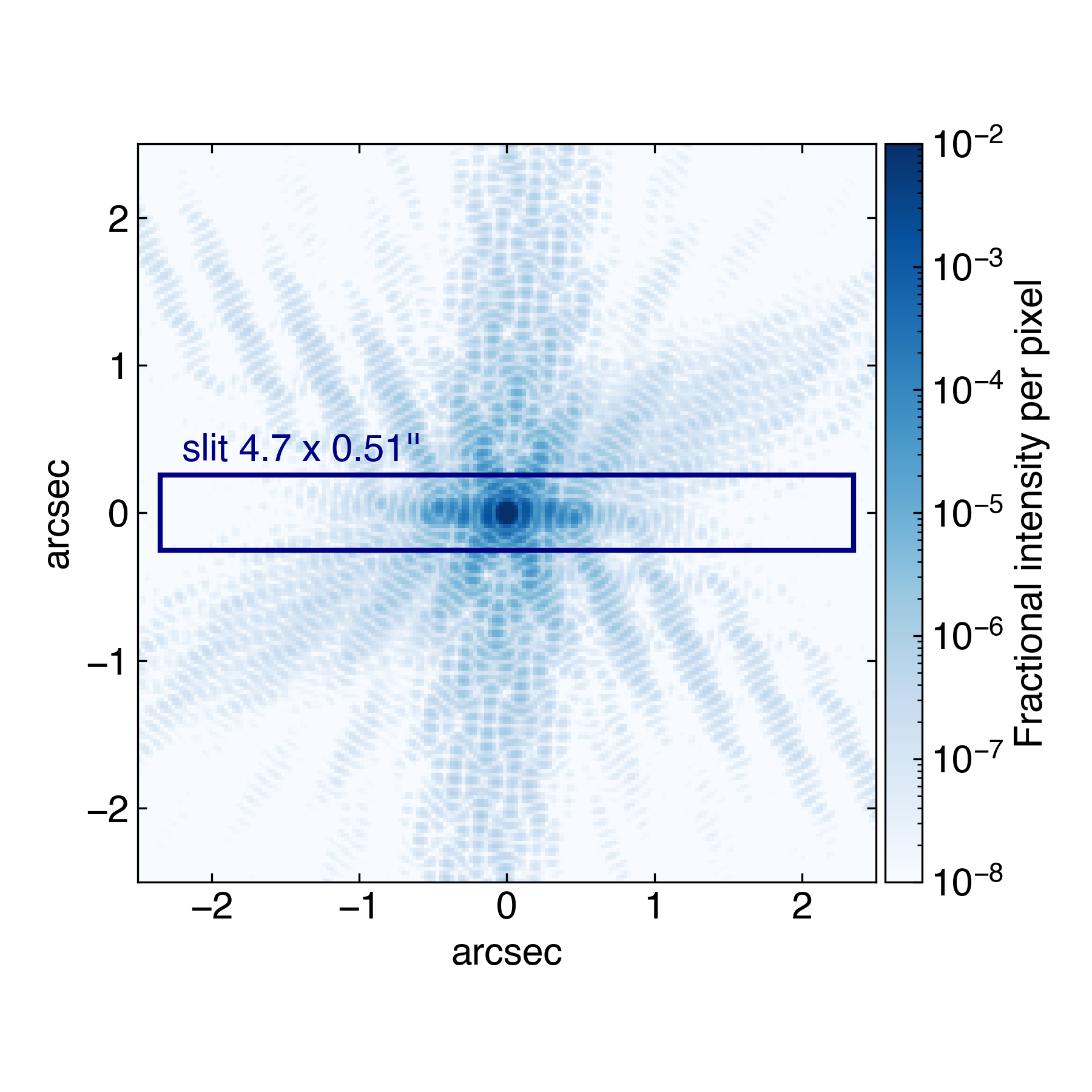}
\caption{Simulated MIRI LRS point spread function at 10~$\mu$m generated with \texttt{stpsf}. The slit is shown for the slit-loss calculation.}
\label{fig:psf}
\end{figure}

The guide-star centroid positions were extracted for the full observation, and the standard deviation of the measured centroid coordinates was computed independently along the detector $x$ and $y$ directions, yielding pointing dispersions of $\sigma_x = 1.29\times10^{-3}$ arcsec and $\sigma_y = 5.92\times10^{-4}$ arcsec. These measured pointing variations were then propagated into simulations of the MIRI/LRS slit using \texttt{stpsf}\cite{oschmann_updated_2014}. A monochromatic PSF was generated at 10~$\mu$m for two configurations: one with the source perfectly centred in the slit and a second with the source displaced by three times the measured pointing dispersion ($3\sigma_x$, $3\sigma_y$), representing a conservative upper limit to the expected pointing excursions during the observation. The total flux transmitted through the slit was computed for both simulations and compared (Fig.~\ref{fig:psf}).

The simulated $3\sigma$ pointing offset changes the total flux transmitted through the slit by 0.004\% (40 ppm). While not strictly negligible, this effect is sufficiently small that it is unlikely to be a dominant source of uncertainty for observations of exoplanet atmospheres. These results indicate that pointing drifts and time-variable slit losses are not expected to limit the performance of the MIRI/LRS slit mode for time-series observations.

\subsection{Slit and slitless background analysis} \label{section:background}

\begin{figure}[htbp]
\centering
\includegraphics[width=0.6\textwidth]{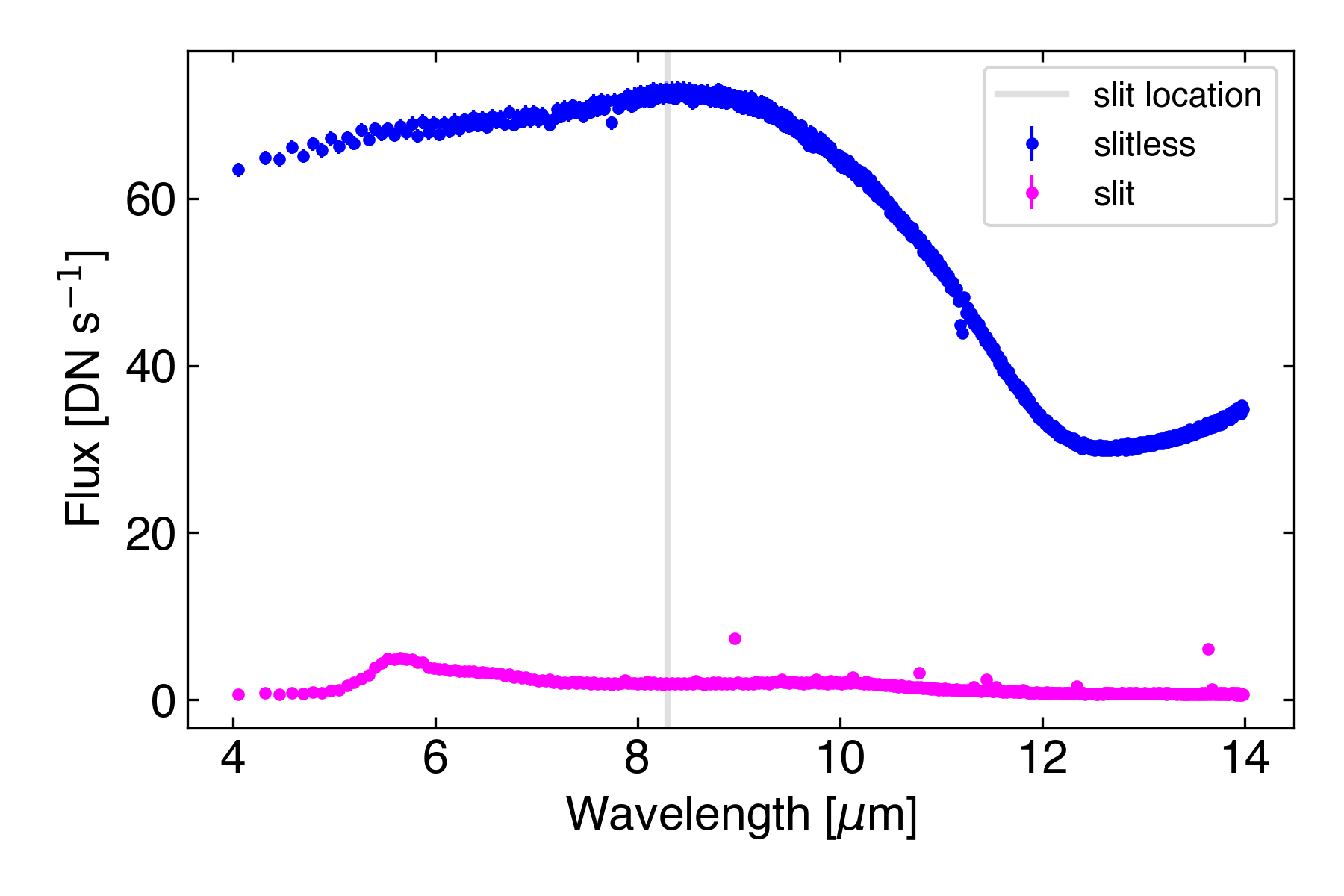}
\caption{Median background values of all integrations for both slit and slitless datasets along the dispersion axis (5--12~$\mu$m), showing a median background reduction of $38\times$ in the slit configuration. The projected centre of the slit location is shown in grey.}
\label{fig:background}
\end{figure}

We compared the background levels measured along the dispersion axis for both the slit and slitless datasets (Fig.~\ref{fig:background}). In the wavelength range covered by MIRI/LRS, the background is dominated by in-field zodiacal emission; thermal self-emission becomes a significant contributor from $\sim$10~$\mu$m\cite{rigby_science_2023}. The background profile is measured along the detector rows for the first integration of both observations. As the detector rows correspond to the spectral direction, higher row numbers map to shorter wavelengths, where the stellar flux is larger. To assess the temporal stability of the background, we also extracted the background time series over the full $\sim$10-hour observations on both the left and right sides of the spectral trace. For the slit observations, the background varies by only $\sim$0.8~DN~s$^{-1}$ over the duration of the observation when accounting for both sides of the trace. The slitless observations show a similarly stable behaviour, with background variations of $\sim$1~DN~s$^{-1}$. These results demonstrate that the background remains nearly constant over time in both observing modes, while the absolute background level is substantially lower in the slit configuration. The slit background is found to be $\sim$38 times lower on average than the slitless background across the 5--12~$\mu$m range. Indeed, the thermal background is considerably larger in slitless observations because each detector pixel receives the integrated background over the full wavelength range, whereas the slit spatially filters the background while simultaneously dispersing it spectrally, resulting in a substantially lower background level. This substantial reduction in background makes the slit mode particularly well suited to time-series observations of faint targets (e.g., $J_{\rm mag} \sim 13$--15), for which background-limited noise can be a dominant source of uncertainty in slitless observations.

\subsection{Estimating wavelength-correlated noise}

\begin{figure}[htbp]
\centering
\includegraphics[width=0.4\textwidth]{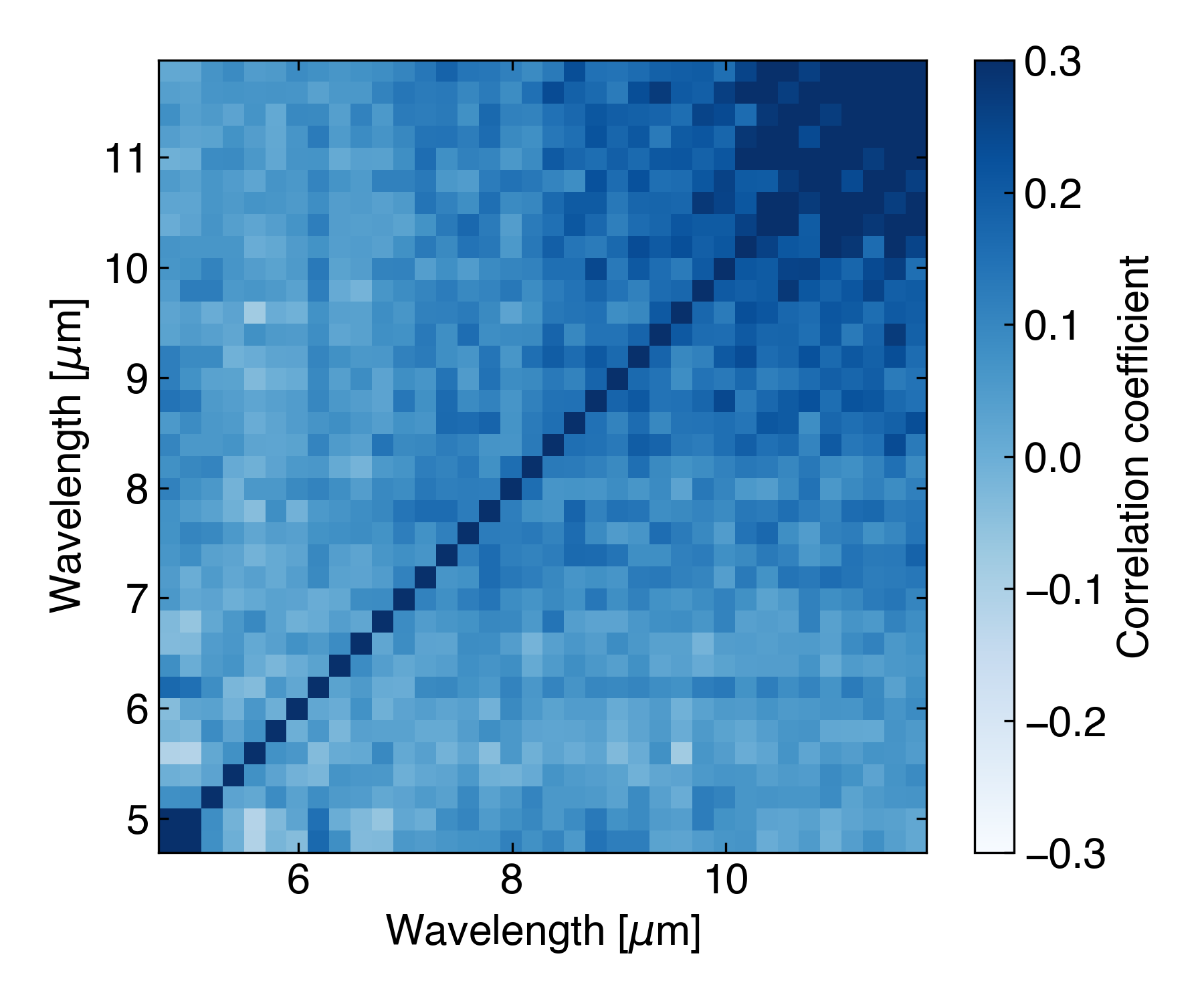}
\includegraphics[width=0.4\textwidth]{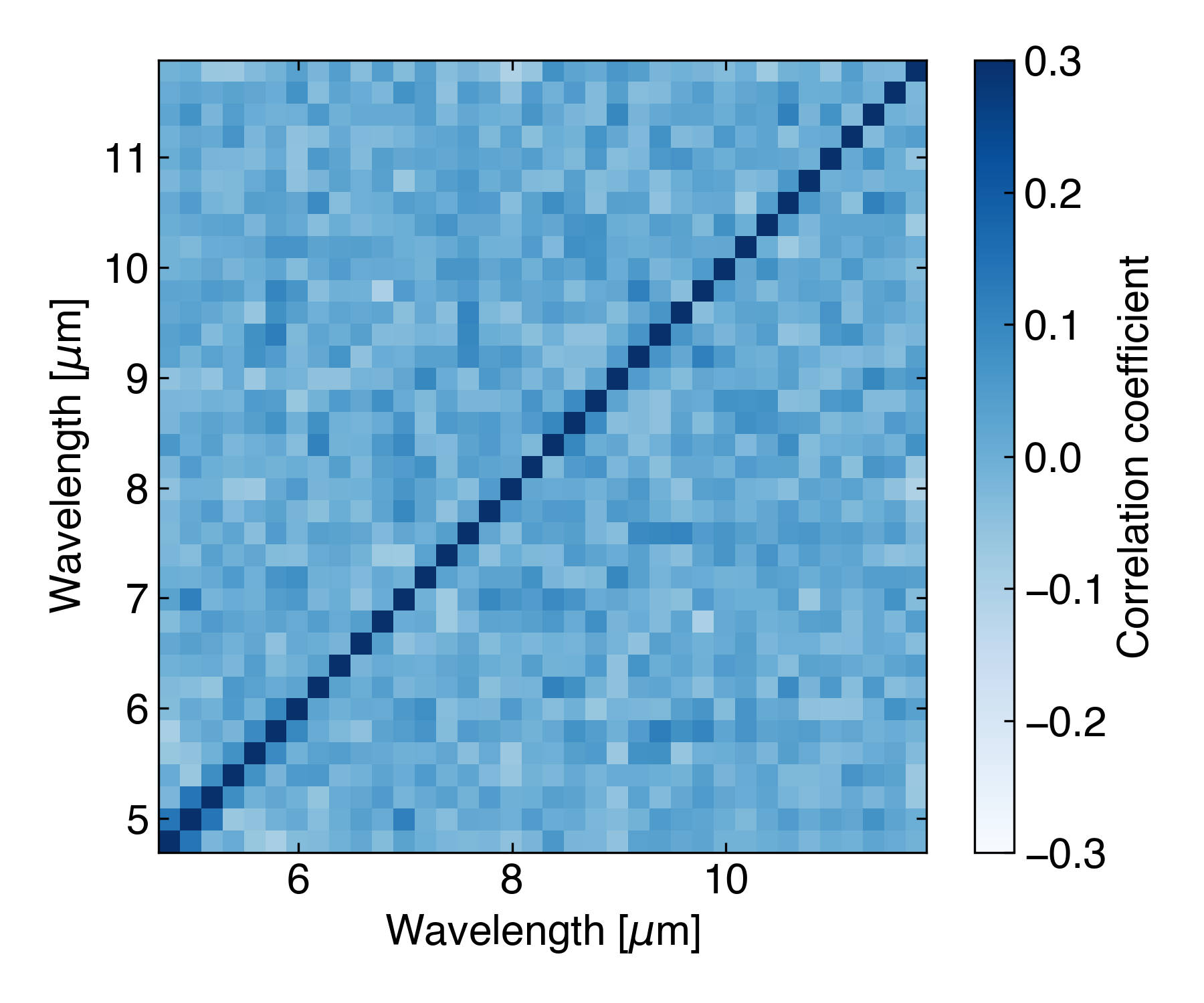}
\caption{Correlation matrices of the residuals from the spectroscopic light-curve fit. \textit{Left:} Matrix for the MIRI/LRS slit observation.  \textit{Right:} Matrix for the MIRI/LRS slitless observation. Off-diagonal elements indicate correlations between residuals at different wavelength channels.}
\label{fig:covariance}
\end{figure}

Following previous analyses of spectroscopic time-series observations \cite{holmberg_exoplanet_2023, radica_awesome_2023}, we investigated the presence of wavelength-dependent correlated noise by computing the correlation matrix of the residuals after fitting the spectroscopic light curves (Fig.~\ref{fig:covariance}). The correlation matrix quantifies the degree to which the residuals at different wavelength channels vary together. In the absence of correlated systematics, the matrix is expected to contain significant signal only along the diagonal, with off-diagonal elements distributed around zero. For the slit observation, the correlation matrix shows a clear increase in positive correlations among the longest wavelength channels, whereas the shorter wavelength channels remain only weakly correlated. In comparison, the slitless observation does not exhibit this long-wavelength correlated structure, with off-diagonal elements remaining close to zero across the spectrum. This behaviour indicates the presence of an additional source of wavelength-correlated noise in the slit observation, primarily affecting wavelengths above $\sim$8~$\mu$m. The origin of this correlated noise is not yet fully understood. 
One possible explanation is that the correlated noise comes from the illumination history of the detector. In the slit observation, the detector pixels containing the science spectrum remain unilluminated before the start of the time-series observation (except for the slit region), whereas in the slitless configuration, they are illuminated beforehand, exposed to previous targets or background and heavily depend on the filter in place, which then can result in systematics, including the shadowed region effect. This difference in illumination history may induce another history-dependent detector response that manifests as wavelength-correlated noise. Further investigation of detector effects, including up-the-ramp non-linearity and persistence in low-flux pixels, will be required to establish the physical origin of these correlations.

\section{CONCLUSIONS}
\label{section:conclusions}

We have demonstrated, for the first time, that time-series observations (TSOs) with the JWST/MIRI LRS slit are feasible and provide photometric and spectroscopic performance suitable for precision exoplanet spectroscopy. Using a $\sim$10-hour transit observation of HAT-P-12b, we showed that telescope pointing jitter produces very low slit-loss variations ($<40$ ppm), while the slit reduces the thermal background by an average factor of 38 compared to the slitless configuration. Independent reductions with two analysis pipelines and a joint fit with archival slitless observations recover consistent transmission spectra and confirm the presence of the atmospheric feature near 7.5~$\mu$m.

Beyond the reduction in background, the slit offers several practical advantages for future time-series observations. By spatially isolating the target, it mitigates contamination from nearby sources and greatly reduces the need for position angle (PA) constraints that are often required in slitless observations to avoid overlapping spectra. Relaxing these constraints will increase scheduling flexibility for JWST observations. Furthermore, because the detector region containing the slit spectrum is largely unilluminated prior to the science observation, the slit configuration provides a more reproducible illumination history than the slitless mode. This may help explain the diversity of detector-settling behaviours reported for slitless TSOs in the literature and could lead to more repeatable instrument systematics between observations.

Our analysis also identifies limitations that would require further investigation. Although the slit reduces the thermal background, the spectroscopic error bars are comparable to those obtained in slitless mode. We attribute this behaviour to the brightness of HAT-P-12, as HAT-P-12 is relatively bright compared to the background. Also, detector effects associated with the short 9-group integrations currently used for slit observations make non-linearity corrections and ramp fitting more challenging than for the slitless configuration whose faster frametime allowed for 150 groups per integration. Indeed, more groups reduce the variance on the fitted count rates. In addition, we detect wavelength-correlated noise at long wavelengths that is unique to the slit observations and whose physical origin remains to be established.

The recent introduction of the SUBSLIT subarray in Cycle~6 (see Kendrew+, 2026, SPIE, submitted) provides a promising avenue to address these limitations by enabling more groups per integration. For very faint objects ($J_{\rm mag}> 25$), the SUBSLIT subarray will both enable longer ramps and limit background asymmetries caused by light coming from the nearly saturated imager\cite{voyer_miri-lrs_2025}. Since the implementation of the slit TSO mode requires only minor modifications to the existing JWST data processing and calibration pipelines as well as observing tools, this new capability can be readily adopted for GO programs from Cycle~7 onward. We anticipate that the MIRI/LRS slit will become the preferred observing mode for faint targets ($J_{\rm mag}\sim13$--15) and for programs requiring robust, repeatable, and contamination-free mid-infrared time-series spectroscopy. 

\acknowledgments
This work is based on observations made with the NASA/ESA/CSA James Webb Space Telescope, obtained under Cycle~3 program PID~6219. The authors heartily thank Jeroen Bouwman for his significant work on this project. The authors thank the JWST/MIRI team and the Space Telescope Science Institute for their support. The authors also thank Margot Courtoux and Marshall Perrin for insightful discussions and help. POL and MV acknowledge funding support from CNES. This project was provided with computing HPC and storage resources by GENCI at TGCC thanks to the grant 2024-15722 and 2025-15722 on the supercomputer Joliot Curie’s SKL and ROME partition.

\bibliography{report} 
\bibliographystyle{spiebib} 

\end{document}